\documentclass[a4paper,12pt]{article}

\usepackage{amssymb,amsmath,amsfonts}
\usepackage[ansinew]{inputenc}
\usepackage{verbatim, graphicx, ulem}
\usepackage[section]{placeins}
\usepackage[english]{babel}
\usepackage{amsmath, amsthm, amssymb}
\usepackage{color}
\usepackage{pgf}
\usepackage{tikz}

\usetikzlibrary{arrows,automata}

\newcommand{\beq}{\begin{equation}}
\newcommand{\eeq}{\end{equation}}
\newcommand{\bdm}{\begin{displaymath}}
\newcommand{\edm}{\end{displaymath}}
\newcommand{\bea}{\begin{eqnarray}}
\newcommand{\eea}{\end{eqnarray}}

\newcommand{\benum}{\begin{enumerate}}
\newcommand{\eenum}{\end{enumerate}}
\newcommand{\bit}{\begin{itemize}}
\newcommand{\eit}{\end{itemize}}
\newcommand{\bdes}{\begin{description}}
\newcommand{\edes}{\end{description}}
\newcommand{\bpic}{\begin{picture}}
\newcommand{\epic}{\end{picture}}
\newcommand{\bc}{\begin{center}}
\newcommand{\ec}{\end{center}}

\newcommand{\bq}{\begin{quote}}
\newcommand{\eq}{\end{quote}}

\begin{document}

\title{Chronic Incompleteness, \\ Final Theory Claims, \\ and the Lack of Free Parameters \\ in String Theory}

\author{Richard Dawid\thanks{Department of Philosophy, Stockholm University,
10691 Stockholm, Sweden.
email: richard.dawid@philosophy.su.se}}

\date{\today}

\maketitle

\begin{abstract} 
String theory has not even come close to a complete formulation after half a century of intense research. On the other hand, a number of features of the theory suggest that the theory, once completed, may be a final theory. It is argued in this chapter that those two conspicuous characteristics of string physics are related to each other. What links them together is the fact that string theory has no dimensionless free parameters at a fundamental level. The paper analyses possible implications of this situation for the long term prospects of theory building in fundamental physics.
\end{abstract}

\section{Introduction}

The present chapter aims at understanding the relation between two remarkable characteristics of string physics. On the one hand, the theory's position within the fabric of physical reasoning as well as some characteristics of the theory itself suggest that string theory, or the theory it would have morphed into once fully developed, might represent a \textit{final theory}: no empirical data can be gathered in this world that would be at variance with the theory's predictions. On the other hand, half a century of intense work on string theory have not resulted in a fully fledged theory. At this point, the prospects that this will happen in the foreseeable future seem, if anything, less promising than at earlier stages of the theory's evolution. 

Each of the described characteristics of string physics raises distinct and substantial philosophical questions. A final theory claim raises the question: what would the research process in fundamental physics look like once a universal final theory has been found? A natural question regarding the chronic incompleteness of string theory is: why is it so difficult to develop string theory into a fully fledged theory? The two questions might actually address two sides of the same medal. They could then be posed in the following more specific way. Are the seemingly insurmountable difficulties to turn string theory into a complete theory conceptually related to its character as a (potential) final theory? Should we therefore understand chronic incompleteness as a core characteristic of a final physical theory? 

Dawid (2006, 2013, 2013a) has suggested that string theory may represent a fundamentally new stage in physical theory building, where theory succession is replaced as the driving principle of conceptual development by the open-ended evolution of the conceptual understanding of a final theory that lacks a time frame for completion. If so, the advent of the final theory prospect would have changed the mode of scientific reasoning without abandoning the `infinite' time horizon associated with the completion of fundamental physics. 

This suggestion was based on two simple observations. i) Chroncical incompleteness and the final theory claim have arisen in parallel in string physics. ii) They seem to reflect generic features of quantum gravity. Without any specific reference to string theory, the current perspective on quantum gravity may be taken to suggest  that the notion of an ongoing succession of theories for ever smaller characteristic distance scales will break down at the Planck scale (see e.g. t'Hooft 1993). Likewise, the chronic problems of attempts to come up with a complete theory of quantum gravity are not confined to the string theory research program. 

Dawid (2006, 2013, 2013a) has not put forward any substantial claim, however, as to how the connection between the final theory status and the chronic incompleteness of string theory can be understood at a conceptual level. 
It is the aim of the present chapter to address this issue and consider some lines of reasoning that may throw more light on the conceptual connection between final theory status and chronic incompleteness. First, I will analyse the ways in which chronic incompleteness may affect our understanding of what a final theory actually amounts to. Second, a highly tentative idea shall be considered as to how finality may be conceptually linked to what seems to be the the predicament of chronic incompleteness faced by physicists working on a universal theory.

What follows is of an explorative nature. It will not present strong conclusions but just test possible ways of thinking about the issues discussed. I do feel however, that those issues are so crucial for understanding the current state of fundamental physics that they should be addressed also from a philosophical perspective, and be it with the highly insufficient means available to the author of this text.

Section 2 lists several ways in which string theory differs from previous stages in fundamental physics. The issues of finality and chronic incompleteness are then discussed specifically in Sections 3 and 4, and related to each other in Sections 5 and 6. Section 7 discusses the role of dualities in the given context, followed by a brief look at the case of M-theory in Section 8.

\section{Five ways in which string physics differs from previous stages in fundamental physics}

In the 19th century, when physical formalization found an ever increasing scope of applications, a canonical understanding of the relation between the observed world and physical theory building emerged. This understanding was based on some core principles. Three of those principles that will be important to our discussion, were:
\vspace{5mm}

I.	A physical theory must be applied to a specified set of phenomena. 

II.	A theory can be adapted to observations by choosing values of dimensionless parameters that either are contained in the theory itself or specify the embedding of the given theory within the physical background needed for understanding the measurements of the described set of phenomena.

III.	Once one has identified a physical context that allows for a formalized account, theories (which might well be insufficient or only partly accurate, to be sure) can be developed (and most often also empirically tested) within a limited time frame. 
\vspace{5mm}

When the basic principles of classical physics were toppled during the physical revolutions of the early decades of the 20th century, this understanding of the role a physical theory was expected to play was complemented by two further principles:

\vspace{5mm}
IV.	A physical theory must be expected to have a limited lifetime as a fundamental theory and to be eventually superseded by a new theory with regard to which it then plays the role of an effective theory.

V.	Advanced physical theories contradict intuitions that are based on everyday world experience and contain parameters that control the deviations from those intuitions.  However, theories that are at variance with intuitions make predictions in line with those intuitions in some limits: special relativity converges towards Galilean physics for velocities very small compared to the speed of light, general relativity approaches predictions of Newtonian gravity in contexts of very small spacetime curvature, and quantum mechanics is consistent with the classical dynamics of objects who interact with the environment and whose action is much larger than $\hbar$.   
\vspace{5mm}

It seems fair to say that principles I - V shape many physicists' expectations regarding the role of physical theory building up to this day. While these principles still work very well in many fields of physics, they came under pressure in fundamental physics once fundamental physics developed a focus on contemplating theories of quantum gravity. This pressure can be felt today whichever approach towards dealing with the problem of quantum gravity is chosen. The present text focuses on string theory, where the inadequacy of the traditional view on physical theory looks particularly far-reaching. In fact, string theory seems flatly incompatible with four of the five principles formulated above and makes an intriguing amendment to the fifth. As opposed to principles I.-V., one may make the following statements about string theory\footnote{Here as throughout the entire paper, the term `string theory', if not specified otherwise, denotes the overall theory that aims at descibing the observed world and is identified by the present knowledge on perturbative superstring theory, duality relations, etc.}

\vspace{5mm}
I'.	String theory is a universal theory.

II'.	String theory has no fundamental dimensionless free parameters (but seems to have a huge number of groundstates).

III'.	String theory is chronically incomplete (and lacks a promising perspective for quantitative empirical testing in the foreseeable future).

IV'.	String theory makes a final theory claim based on a minimal length scale.		

V'.	String theory has several classical limits related to each other by duality relations.

\vspace{5mm}	
Given that the five described significant shifts with regard to the role of theory building arise in the same theory, it seems natural to suspect that they are somewhat related to each other. In the following, I will aim to turn that suspicion into something slightly more concrete.

\section{Finality}

A plausible starting point for finding connections between the five statements on string theory is point IV'. T-duality implies that the string scale amounts to a minimal length scale in string theory. Any statement on a distance smaller than the string scale can be formulated, in the T-dual picture, as a statement about a distance larger than the string length $l$. This fact may be taken to indicate that string theory, if viable\footnote{Throughout this chapter, I will talk about a theory's viability rather than its truth. A theory is viable in a given regime (e.g. up to a given energy scale), if it is in agreement with all the data than can be collected in that regime (e.g. up to that energy scale). Talking about a theory's truth raises difficult philosophical questions that are not relevant to the issues addressed in this chapter. A more extensive argument in favor of the concept of viability can be found in Dawid (forthcoming).} up to its own characteristic scale, correctly represents physics at all scales and therefore constitutes a final theory (Witten 1996). 

A final theory claim that is extracted from a specific theory may be suspected to be begging the question. If string theory were viable only as an effective theory of a more fundamental theory, smaller distance scales could carry new information based on that more fundamental theory. If so, the more fundamental theory would break the final theory claim that had been developed based on what turned out to be its effective theory. A final theory claim based on the allegedly final theory's own implications therefore may seem to deflate to the trivial claim: if the theory is absolutely true, it is actually absolutely true. Dawid (2013, 2013a) argues that, despite the described issue, final theory claims can retain argumentative power in conjunction with arguments of limitations to scientific underdetermination. 

In the present discussion, we will therefore assume that the final theory claim stated in point IV', though obviously not conclusive, does carry argumentative weight. If so, points I' and II' and IV' form a remarkable set of characteristics of string theory that suggest finality. Universality (point I') removes the need for theory succession on the path to further unification and thereby frames string theory as a plausible endpoint of theory succession. The lack of fundamental dimensionless free parameters (point II') removes the need for theory succession on the path towards explaining specific parameter values (since all effective parameter values must result from the dynamics of the full theory, where no parameter values can be chosen at will) and thereby reinforces the former claim. Finally, the explicit final theory claim (point IV') provides a conceptual reason within the theory's theoretical framework for assuming the finality of string theory. In conjunction, the three points form a consistent and mutually reinforcing set of arguments for string theory's status as a final theory. Viewing the issue the opposite way, Points I', II' and IV' are per se highly unusual properties that may be expected to arise in conjunction in a final theory.

\section{Chronic Incompleteness}

	   If it were just for the three discussed new aspects of string theory, one would thus face the stunning perspective of an imminent end of fundamental physics: once the final theory would have been fully formulated, no new fundamental theories should be expected to emerge. Physics would be reduced to the menial work of developing theories in more specific low energy contexts and understanding their reductive connections to string theory.
	
	String theory comes with another substantial shift, however, that renders the above understanding inadequate. 50 years after the birth of string theory (Veneziano 1968) and 44 years after string theory has been first proposed as a universal theory of all interactions (Scherk and Schwarz 1974), the theory still lacks a plausible perspective for a full formulation in the foreseeable future. The theory has known periods of considerable optimism. At least twice during its evolution, in the years after the consistency of a superstring action had been shown by Green and Schwarz (1984), and in the years after the discovery of the web of dualities (Witten 1995, Townsend 1995, Polchinski 1996)  and AdS/CFT correspondence (Maldacena 1998), string theorists were hopeful that a full formulation of the theory was attainable by developing further and making full use of  the concepts and tools available at the time. But each time overwhelming obstacles to achieving that goal soon surfaced as if they were put up by magic hands defending the ultimate secrets of physics. The problems associated with completing string theory in each case turned out to transcend what string theorists at the given point had taken them to be.
	
	Today it seems clear that, even if the final theory claim regarding string theory were true, this fact would not translate into a prospect of the imminent end of fundamental physics. Rather than bringing the time horizon for the completion of fundamental physics from virtual infinity to somewhere within our lifetime, string theory's final theory claim seems to be associated with an extension of the time horizon for the completion of this particular theory that may, once again, virtually reach towards infinity. 

\section{Chronic Incompleteness and the Lack of Free Parameters}
	
Why is it so immensely difficult to turn string theory into a complete theory? One important reason arguably has to do with one of the three points associated with string theory's finality claim: point II', which states the the lack of dimensionless free parameters at a fundamental level, creates a new kind of problem for a theory's completion.

In order to get a better grasp of this point, one needs to start with a look at perturbation theory. 

In QFT, which is understood to be the theoretical framework for calculating low energy effective theories of string theory, calculations of scattering processes are based on a perturbative expansion in the coupling constant. Small coupling constants allow for increasingly accurate results of calculations up to a few orders in perturbation theory. 

It is one important aspect of quantum field theory that the strong coupling regime is directly related to the deep quantum regime (see e.g. Polchinski 2017). Higher orders in the coupling constants correspond to higher orders in $\hbar$. A weak coupling limit therefore corresponds to a small $\hbar$ limit and represents a physical situation where elementary objects can be localized fairly well. Physics then is in a near-classical regime. 
 
Quantum mechanics frames measurements in terms of confronting a quantum system with a quasi-classical measuring apparatus. The entire setup of a quantum theory is therefore developed from the perspective of a classical limit. 

As long as the theory is in a regime where the coupling constant is small, perturbation theory works. Once one aims to describe a strong coupling regime, perturbation theory breaks down and one needs to rely on non-perturbative methods and auxiliary techniques such as lattice theory. Such a situation corresponds to a fully quantum regime where elementary objects are not well localized.

In the standard model of particle physics, the electroweak sector lives in a weak coupling regime and therefore allows for perturbative calculations. QCD, at low energies, to the contrary, is strongly coupled and requires non-perturbative techniques. 

String theory has been developed based on perturbation theory in the same mold as perturbation theory for QFT. It describes the propagation and scattering of oscillating quantum strings on a spatiotemporal background.  
It is clear, however, that this is an insufficient approach for spelling out the full theory. The spatiotemporal background is itself generated by stringy dynamics. 
And string theoretical scenarios may be placed in either strong or weak coupling regimes. In other words, it is clear that a full understanding of string theory must reach out beyond the perturbative regime. 

Duality relations constitute the most powerful tools in a string theoretical context to reach out beyond the perturbative regime. The power of dualities was first clearly understood in the mid 1990s, when Witten (1995) conjectured that the five types of superstring theory plus a sixth enigmatic 11-dimensional M-theory were related to each other by S- and T-duality relations that turned them into one different representation of the same theory. 

Exact dualities relate different theories or models to each other that are empirically equivalent.\footnote{The analysis of this chapter is based on the understanding that duality relations are exact. Recently, the possibility of an inexact notion of dualities has been emphasised in a philosophical context by De Haro (2018).} The isomorphy between the dual models is established based on a `translation manual' that indicates which parameters in one theory correspond to which parameters in the dual theory and how the values of those parameters in one theory are related to the values of the dual parameter in the dual model. 

Duality transformations are particularily helpful for reaching out towards the non-perturbative regime because they tend to invert parameter values and thereby relate one theory in a deep quantum regime to another one in a near classical regime. S-duality relates a theory with a strong string coupling $g$ to a theory with a weak string coupling $1/g$. T-duality relates a theory with small compactification radius $r$ to a theory with a large compactification radius $1/r$. In supersting theory, different types of superstring theory are connected that way. For example, S-duality relates Type I superstring theory with a given string coupling g to SO(32) heterotic string theory with string coupling $1/g$, which in turn has as its T-dual $E_{8} {\rm x} E_{8}$ heterotic string theory with an inverted compactified dimension. 
A strongly coupled type I string theory can therefore be represented as a weakly coupled SO(32) heterotic string theory. And if that theory contains a small compact dimension, it can be represented as a $E_{8} {\rm x} E_{8}$ heterotic string theory with a large compactification radius. The web of dualities therefore extends the reach of string theory by establishing that large parts of what seemed to lie beyond the reach of perturbation theory is accessible by perturbation theory in a dual representation.

But string theory faces a problem that is substantially more serious than the problem faced by a strongly coupled gauge theory. The additional problem is related to the fact that string theory is a theory without fundamental dimensionless free parameters.

In physics, it is important to distinguish between dimensionful and dimensionless parameters.\footnote{(For an instructive analysis of the role of free parameters in physics, see Duff 2015.)} The absolute value of a dimensionful parameter is a matter of choosing a physical unit. Per se, it is irrelevant for fitting theory to experiment. The value of a dimensionless parameter, to the contrary, cannot be altered by choosing different physical units. It therefore tells something significant about our set of observations (if fixed based on empirical data) or about a theory's empirical implications (if extracted from the theory itself). 
Physical theories normally involve dimensionless free parameters that can be tuned in order to fit the quantitative specifics of the empirical evidence. In many cases, for example in quantum theories that  involve dimensionless coupling constants, dimensionless free parameters are an element of the theory itself. In other cases, for example in Newtonian gravity,
the theory's embedding within our observed world requires a comparison of a dimensionful constant (such as the gravitational constant) with constants that characterize the empirical setup used to test that theory. Free dimensionless parameters then arise in the context of this comparison. In both cases, the free parameters involved play a crucial role for connecting the theory to observation.

String theory is a fully universal theory that gives a joint description of all fundamental phenomena. Measuring a fully universal theory cannot depend on paramater values that are not represented in the theory itself. Dimensionless free parameters thus cannot be generated by embedding a fully universal theory in a wider framework. The issue of free dimensionless parameters is reduced to internal characteristics of the theory itself. 

String theory has one dimensionful free parameter, the string length $l$. But the theory has no dimensionless free parameter. In the context of perturbative string theory, this fact can be understood by looking at the geometric structure of Feynman diagrams. While Feynman diagrams for pointlike particles have pointlike interaction nodes where dimensionless coupling constants can be inserted, the corresponding stringy Feynman diagram is represented by a two dimensional surface with non-trivial topology. No coupling constants can be inserted at any point in those diagrams. The string coupling enters as the value of an oszillation mode of the string, the dilaton. 

The lack of dimensionless free parameters in string theory is crucial for our analysis because dimensionless free parameters play an important role in the specification of classical limits. Describing a near-classical limit in QFT corresponds to choosing a small value of a dimensionless coupling constant. Describing a small curvature scenario corresponds to choosing a small value of the ratio between the characteristic length scale of the dynamics described and the radius  of  spacetime curvature. In the case of theories that involve free parameters, placing the theory close to or far from a classical limit can be controlled by choosing a value of a free dimensionless paramter. In string theory, nothing of this sort can be done. The fundamental dynamics of string theory therefore cannot be exempified by choosing a near-classical limit. Whether or not the theory has solutions that lie close to a classical limit is decided by the fundamental dynamics itself. 

To be sure, parameter values such as the string coupling, the radii of compact dimensions or background curvature terms can be specified in effective models that emerge from the theory's fundamental dynamics and correspond to a ground state of the theory. Those effective parameter values can then be close to or far from a classical limit. Calculating the fundamental dynamics itself, however, cannot rely on specifying a parameter value that controls whether physics is in a near classical or in a deep quantum regime. Only once effective parameter values have been extracted as a solution of the fundamental dynamics, does the categorization in terms of near classical or deep quantum make sense. Curvature terms, compactification radii or string coupling constants can only be specified in the ground state of the theory once the fundamental dynamics of the theory has already played out. 

String theory thus finds itself in the following peculiar situation. It is defined, as it stands, at a perturbative level in terms of strings moving through background space. The core parameters that characterise the theory's state are spatio-temporal and allow for calculations in a near classical limit where interaction is weak, objects are localized and curvature is low. The propagation of individual strings only amounts to a plausible intuitive representation of the dynamics in a near classical limit.  But the fundamental theory, lacking dimensionless free parameters, cannot be related to any such limit before calculation. 

Framing the theory in terms of parameters that are selected in order to work well in the near classical limit therefore looks like an awkward choice when it comes to describing and calculating string theory's fundamental dynamics. Still, all physical parameters deployed for describing string theory or any other physical theory are of that kind (with one possible exception we will return to a little later). These parameters seem capable of determining some contours of the fundamental theory with the help of duality arguments and consistency requirements. But they have not proved capable of providing a workable basis for calculating the theory's fundamental dynamics. 

\section{Finality in the face of chronic incompleteness}
 
Let us, for the moment, presume that string theory is a final theory in the sense that i) there is a string theory ground state that fully represents phenomenology at energies below the Planck scale and ii) no observation that can be made in principle would lead to the rejection of string theory. If string theory is a final theory in this sense, the pair of facts stated at the end of the previous paragraph may have at least three possible explanations.

\vspace{5mm}

A. Even in principle, there exists no mathematical scheme that is empirically equivalent to string theory and generates quantitative results that specify the fundamental dynamics of the theory. In that case, the fundamental theory is conceptually incomplete by its very nature. It has no fundamental dynamics and no set of solutions that can be deduced from its first principles. The fundamental theory merely serves as a conceptual shell that embeds low energy descriptions (ground states of the theory) consistent with the principles encoded in the fundamental theory. Those low energy descriptions contain specified parameter values and do generate quantitative results. But there is no way to establish from first principles how probable specific ground states of the system are. 


B.	Mathematically, the fundamental theory does have well defined solutions. There is a fact of the matter as to whether or not a low energy state constitutes a solution of the fundamental theory and as to how probable that state is. However, there exists no algorithm that can establish those quantitative results based on calculations that are substantially less complex than the full description of the entire dynamics of the universe. Therefore, those calculations will never be carried out by physicists. 


C.	There is a mathematical scheme that allows for calculating the dynamics of the fundamental theory with reasonable effort. Physicists just have not found it yet.

\vspace{5mm}

Let us look a little closer at those three possibilities.

Alternative A would imply a fundamental break with the way science has understood the role of mathematics in representing observations. In particular, it would raise fairly intricate philosophical questions regarding the conception of a real world and its relation to its mathematical description. 

If string theory were indeed viable and A were true, the dynamics of the world could not even probabilistically be extracted from a formal fundamental theory. A fundamental theory is here understood to be constituted by a formal structure that is defined based on a finite number of mathematical posits. In this scenario, quantitative treatment and the extraction of quantitative predictions could be achieved at the level string theory groundstates. That level of description could not be understood as fundamental, however, since its consistent conceptualization relies on the representation of that dynamics in terms of groundstates of a fundamental theory. The corresponding theory could be inferred in its basic characteristics but could not even in principle be expanded into a conceptual scheme that allowed for a specification of the fundmanetal dynamics and the extraction of quantitative predictions.   

A situation of that kind would allow for various philosophical interpretations. From a strictly empiricist point of view, it could simply be characterized in terms of limitations to the predictive power of scientific theory building. The peculiar status of the fundamental theory, which is at the same time implied by the requirement of having a consistent notion of string theory groundstates but also fundamentally incapable of being turned into a predictive scheme, is not reflected by a specific philosophical conceptualization from such an empiricist perspective.

Viewed from the perspective of scientific realism, to the contrary, alternative A raises fundamental new questions regarding the relation between reality and physical theory. Those questions can once again be addressed in a number of different ways.

First, one might conclude that there is a part of reality that just cannot be represented by mathematics. But how, if not based on mathematics, should one conceive of that part of reality? Given the full dominance of mathematical conceptualization and the radical dissolution of any intuition-based concepts at the most fundamental levels of physical description, prospects for developing a convincing conceptual basis for any such understanding seem very limited. 
If there is any philosophical message one may consistently draw from the evolution of physics throughout the 20th century, it hints in exactly the opposite direction. The century started with the emergence of a general consensus that human intuitions regarding material objects and their movement through space could be deployed in the context of microphysics in order to develop a deeper - atomist - understanding of the world. The further the century progressed, however, the more conspicuous it became that each further conceptual step in the development of fundamental physics amounted to jettisoning core elements of that intuitive perspective. 
Most ideas on how to join general relativity and quantum physics seem to imply that, as a last step in this decay process of intuitive concepts in fundamental physics, space and time themselves need to be abandoned as fundamental concepts. Having lost all guidelines that had been based on human intuitions regarding physical objects, mathematics seemed the only remaining basis for adequately expressing the fundamental physical characteristics of the world. Under those circumstances, the suggestion to aim at some non-mathematical perspective on fundamental aspects of the world seems hopelessly backwards-leaning.  
   
In this light, it might look more promising to view alternative A from a very different perspective. Reality, on that alternative reading, does not reach beyond what can be grasped based on mathematical formalization but, quite to the contrary, should be regarded as being confined to those levels of mathematical description that allow for a quantitative analysis. If so, there would be no reality corresponding to the level of fundamental principles for the very reason that those principles don't imply a fundamental dynamics. The fundamental theory would assume the peculiar role of providing the closure of physics that is needed for consistency reasons without representing the real world. 

A perspective that avoids this irritating idea would be to insist on the principle that there is a direct correspondence between the mathematical structure of a true physical theory and the real world. Structural reality could then be attributed to the fundamental level. At that level, however, it would play out in a way very different from what we expect from a physical theory. Only at the level of effective theories that allow for a quantitatively specified dynamics would reality resemble a physical system in the sense we are accustomed to.
 
Whatever the philosophical interpretation however, alternative A amounts to introducing a strictly non-reductionist element into the fabric of physical theory building: neither at a conceptual nor at an ontological level would it be possible to reduce the phenomenology describable by our effective theory to the principles of the fundamental theory. 

Alternative B) has the same practical implications as A) but remains closer to known philosophical territory. Mathematics could remain in place as a complete representation of the dynamics of the world at the fundamental level. Reductionism and full consistency could be upheld as general ontic principles of physics. There would be a set of fundamental equations that fully specify the state of the world (or, at any rate, the probability of the state we find ourselves in). Alternative B would just imply that no manageable reduction of complexity regarding the fundamental dynamics of the world can be achieved that leads to the extraction of parameter values of effective theories based on any method of approximation. As briefly mentioned above, physics knows contexts that are sufficiently chaotic to block any serious attempt at calculating specific outcomes of the dynamics.  Just think of the exact point of landing of a leaf carried away by autumn winds.\footnote{Some philosophers of science (see e.g. Cartwright 1983) have drawn far-reaching anti-reductionist conclusions from this fact.} In the given case, one might suspect an even more desperate situation. Given the lack of free parameters and classical initial conditions, the case we face might not even allow for choosing simple initial conditions that do allow for calculations of the fundamental dynamics in a toy model. 

This outcome would retain the traditional philosophical take on the role of mathematics in physical theorizing. But it would not provide a physical perspective for understanding the fundamental basis of the dynamics we observe at low energies.

Alternative C) is what physicists are hoping for. Are there any good reasons for assuming the existence of a full-fledged and calculable fundamental theory of all interactions? One reason for feeling inclined to give a positive answer to this question may lie in the history of science. The conviction that physical phenomena can be described by a fully calculable theory  even if they seem to defy any such description at first glance has been the main driving force in physics ever since the time of Galilei. The striking successes of physical theory building during four centuries bear witness to the fact that this conviction has often been vindicated. 

	Two caveats should be pointed at, however. First, as mentioned above, not all physical problems are assumed to have a calculable solution. In this light, the conviction described above is in need of qualification. Scientists first make up their minds, based on their experience, as to whether or not a given physical problem can be expected to have a calculable solution. The physics community's track record of predicting what would eventually be calculable obviously is not flawless but seems fairly good on average.  In light of this track record, the physicists' expectation that a given physical problem has a calculable answer can become very strong. Such strong expectations do constitute a main driving force behind their work. 
	
A theory of quantum gravity, or string theory for that matter, at first glance looks like the kind of theory that should be possible to develop and eventually to calculate. But, and this brings the second caveat into play, string theory and the conceptual context within which it is developed is in a number of ways substantially different from anything physicists have witnessed up to this point. Therefore, it is far from clear whether prevalent physical intuitions as to which kinds of questions can be expected to have a fully calculable theoretical answer are applicable in this case. It seems difficult to rule out that what seems to be a question that finds a fully calculable theoretical answer in fact rather resembles the case of the leaf carried by autumn winds and just defies calculation. This would bring us back to alternative B. 

One specific difference between string theory and QFT brings alternative A into play as well. In QFT, the perturbative approach is a tool for extracting quantitative predictions from the theory. But QFT theory can be formulated as a full fledged theory (albeit not with the rigor algebraic quantum field theorists would want). In the case of string theory, the perturbative approach was all there was in the early days of the theory. No formulation of the theory was known that did not rely on the perturbative approach. Though it was expected that the perturbative formulation of the theory should in a cogent way designate a coherent full theory of strings, it was by no means clear that such a theory existed. 

The conspicuous abundance of string dualities is by many taken to offer a physical reason to believe in the existence of a calculable fundamental theory. The conjecture of, let us say, an exact S-duality relation, implies that deeply non-perturbative large coupling solutions exist for each of the types of string theory that are related to each other by the duality. This may be read as a stepping stone towards a general fundamental formulation of string theory .  

In the late 1990s there was much confidence among string theorists that dualities could be developed into a tool for the full calculation of string theory's dynamics\footnote{A good example of the optimistic spirit at the end of the 20th century is (Greene 1999).}. These hopes have turned out to have been premature. Though dualities have led to many new insights into string theory, it is now generally acknowledged that further substantially new methods would be needed in order to fulfil their early promises. Dualities, in this light, may still be viewed as an indicator that full calculability is a serious possibility but, at this point, cannot establish that a fully calculable theory exists.

Given the power but also the limitations of duality relations with regard to finding a fully calculable string theory, how should one understand the role of dualities in this search on a more general basis? 

The next Section will analyse this issue by looking at the various kinds of dualities that have been discovered so far in the context of string theory.

\section{Types of duality in string physics}

The extent to which parameters change from a model to its dual varies depending on the type of duality and the context within which it plays out. The narrowest instantiation of a duality relates parameter values to inverted values of the same parameter in the same theory. This form of duality is called self duality. An example in string physics is the  S- self duality of Type IIB superstring theory: type IIB string theory with a string coupling $g$ is dual to itself with a string coupling $1/g$. 
Another example is the T-duality of Bosonic string theory: a Bosonic string theory with compactified radius $R$ is dual to itself with compactified radius $l^2/R$ (where $l$ is the string length).

In those cases, observable characteristics of individual objects may appear as different characteristics in the dual description. For example, the winding number of a string corresponds to the transversal momentum of the string in the T-dual description.

Dualities can also relate values of a given parameter to the inverted values of that parameter in a different theory. We have already encountered the example of an S-duality that relates Type I superstring theory with a given string coupling g to SO(32) heterotic string theory with string coupling $1/g$, which in turn has as its T-dual $E_{8} {\rm x} E_{8}$ heterotic string theory with an inverted compactified dimension. 

Duality relations can reach out even farther, however, and relate very different parameters in very different theories to each other. 
An example of this kind is AdS/CFT duality. To state one specific examplification of AdS/CFT, a type IIB superstring theory on $AdS_{5} {\rm x} S_{5}$ space is related to a four dimensional N=4 supersymmetric SU(n) Yang Mills theory (which is a conformal field theory). The duality relates the curvature of AdS (in units of the string length) to the 't Hooft parameter  $(g_{YM}n)^{1/4}$ in the dual conformal field theory, where $g_{YM}$ is the coupling constant of the Yang Mills theory. In other words, a superstring theory is conjectured to be dual to a theory that is conceptually so different that parameters that characterize the first theory (such as spacetime curvature) don't even show up as such in the dual. The duality relation therefore links values of that parameter to values of an entirely different parameter in the dual theory.

One characteristic is common to all exemplifications of duality mentioned so far: each of the parameter values linked by duality relations controls a classical or low curvature limit of the respective theory. Moving away from a classical limit in one theory corresponds to moving closer to a classical limit in its dual. We have discussed this feature already in the S- and T-duality cases. S-duality relates a coupling constant to its inverse. Moving to higher -- and thereby more quantum -- values of the coupling in one theory therefore means moving to smaller -- and therefore more classical -- values in the dual. Similarly, if one decreases the compactification radius of a theory -- and thereby moves towards a situation where the classical concept of localization on that dimension becomes less meaningful -- one increases the compactification radius of the T-dual, thereby moving closer to a classical intuitive understanding of that dimension. 

In the AdS/CFT case, this principle still applies. String theory on AdS can be related to an effective scenario of localized massive objects moving through space only if the curvature radius is much larger than the string length. CFT, on the other hand, has the characteristics of a weakly coupled theory only if the 't Hooft parameter is small. And a small 't Hooft parameter corresponds to a small curvature radius. Therefore, each of the dual theories has a classical limit that is controlled by parameters that are linked to each other by the duality relation. Moving closer to the classical limit in one theory amounts to moving farther away from the classical limit in the dual. 

As we have seen,  it is exactly this typical characteristic of duality relations encountered in the context of string physics that has proved so helpful for reaching out beyond the perturbative regime. It can allow for calculations of physical situations that don't look near classical in terms of one theory but do look near-classical in its dual. However, for the reasons discussed in Section 5, the described characteristic is not helpful for calculating the fundamental dynamics of a theory without free parameters. The core obstacle to any such calculation is the perturbative approach whose reliability is confined to specific parameter values while the fundamental dynamics must be calculated before any such parameter values have been fixed. 

A further extension of the reach of duality relations might, in principle, overcome that obstacle: a duality that relates one parameter in the initial theory that controls a classical limit to a different parameter in the dual theory that does not lead towards a classical limit at all.  Obviously such a theory would encode the classical limit of its dual theory in a similar way as, to take one example, a strongly coupled Type I superstring theory encodes the weak coupling limit of SO(32) heterotic string theory . But a theory of the described kind would not contain elementary objects which, for specific parameters values, can be localized and behave in a near classical way. 
Such a theory, if it exists, cannot restrict calculability to a near classical limit because there is no such limit close to which perturbation theory works. So either the theory is nowhere calculable (which means that there is no regime where it can make any quantitative predictions) or it must be calculable away from any classical limit. That is, it had to be calculable without perturbing around a classical solution.
In that case, the theory might offer a promising framework for a full calculation of the fundamental dynamics. 

If such a theory exists, it would be immensely difficult to find. New theories and models are normally found by first thinking about ways in which one can consistently move away from a classical limit. Once a new general theory has been developed, specific realizations of that theory are found by trying to reproduce empirical data, solving conceptual problems faced by specific model building, and scanning the options for consistent theory and model building within the framework provided by the general theory. To give an example, quantum field theory provided a general framework within which specific theories and models, such as the standard model, supersymmetric models, etc.  were then developed. 

Dualities are normally discovered by relating two theories to each other that had already been found along the lines just decribed but had been taken to be substantially different due to their different classical limits. When Montonen and Olive (1977) conjectured a duality between a theory with magnetic solitonic monopoles and one with electric ones, the surprising point was the duality structure that related weak to strong coupling limits rather than the point that electric solitons could be built just like magnetic ones. When S- and T-duality were understood to link the five types of superstring theory, all five theories had already been spelled out before. The same goes for the two sides of AdS/CFT duality. There is no well-established method to construct a new theory based on conjecturing a duality relation between a known and an unknown theory and constructing the unknown theory just by applying the duality. 

So let us now assume that there is a duality relation that connects a theory with a classical limit to a theory that does not relate classical limits to specific values of its parameters. The latter theory cannot be found based on the traditional heuristics of theory development that are based on thinking in terms of near classical limits. It will also be very difficult to understand that the duality exists at all, given that, as pointed out above, duality relations typically are discovered based on prior knowledge about the two theories that turn out to be dual.
Neither the theory nor the corresponding duality relation would be likely to be discovered in such a scenario.

 One might therefore imagine the following scenario. There exist, in principle, two kinds of theories about fundamental physics. There are those which can provide quantitative calculations of  near classical limits based on perturbation theory. Such theories are adequate for effective theories whose parameter values do sit close to a classical limit and can be discovered and developed by thinking about them in terms of a near-classical limit. But they are not adequate for calculating the dynamics of the final theory (that is, a theory without free parameters). 

Then, there are those theories that are not represented by any parameters that lead towards a classical limit. Therefore they don't have parameter values that are preferable for calculations. 
To the extent those theories are calculable at all, they may be expected to allow for the calculation of the entire parameter space and, in this light look like adequate theories for representing the dynamics of a fundamental theory without free parameters. But they are hidden in a conceptual realm that cannot be accessed by following near-classical intuitions.

\section{The Case of M-Theory}

There is one important theory in the context of string physics that may represent an example of the described scenario: M-theory. 

Type IIA string theory with a given string coupling $g$ is conjectured to be dual to an eleven dimensional theory of membranes and five-branes (M-theory) with a compactified 11th dimension. $E_{8} {\rm x}E_{8}$ heterotic string theory is conjectured to be dual to an 11-dimensional theory of membranes and five branes where the 11th dimension is bounded by two nine-branes, each of them carrying one set of $E_{8}$ gauge fields. The size of the eleventh dimension (and the membrane extending through that eleventh dimension) in each case corresponds to the size of the string coupling. The strong coupling limits of the string theories in both cases corresponds to a dual limit of M-theory with an infinitely extended 11th dimension.  
 
The relation between M-theory and the two types of string theory dual to it is of a peculiar kind. In the low coupling limit, the radius of the 11th dimension of M-theory goes to zero. The theory dual to M-theory therefore does have a classical limit of the kind known from other duality relations.  But M-theory with a large 11th dimension does not amount to a new classical limit. It corresponds to full uncompactified M-theory. M-theory does have a low energy limit (covering energy scales small compared to the string length) where 11 dimensional supergravity serves as its effective theory. In this sense, one can reach a new classical limit that amounts to the classical limit of a quantum field theory of point-like objects in 11 dimensions. But that limit is not taken in a genuine M-theory description. 11 dimensional supergravity knows nothing about two and five branes. 

Therefore, the relation between M-theory and Type IIA or $E_{8}\rm{x}E_{8}$ heterotic string theory is quite different from all other known duality relations. By lacking a classical limit, it resembles the scenario that was described before as a scenario that may generate the prospect of a fully calculable theory. 

The lack of a perturbative regime for M-theory also constitutes on core obstacle to formulating M-theory even to the extent this has been achieved for 10-dimensional superstring theories. The pertrubative `entry point' that is available in the case of superstring theories is not available in the case of M-theory. More than 20 years after it was conjectured, it is thus still unclear how to turn a theory of two- and five-branes in 11 dimensions into a consistent quantum theory. The concepts of membranes and fivebranes provide a good heuristics for understanding the nature of the duality relation. In the absence of a genuinely `membrany' regime that allows for a perturbative description, it seems doubtful, however, whether membranes and fivebranes provide a promising framework for fully formulating and calculating the theory. A lot of energy has been spent on finding better parameters for representing the theory. Matrix theory (Banks et al. 1997) has been an influential candidate but has faced problems of its own. In other words, understanding M-theory suffers from the fact that the heuristics of theory development cannot be guided by looking at a classical limit of the theory itself.

\section{Conclusion} 

The present chapter has argued that the lack of free parameters of string theory plays a crucial role in conceptually connecting two striking features of string physics: string theory's final theory claim and its chronic incompleteness. For several reasons, a lack of free parameters may be taken to be a natural feature of a final and universal theory.
But the lack of free parameters removes the possibility of calculating the theory's dynamics starting from a classical limit based on a perturbative expansion. 
Full access to a theory without free parameters thus might be expected to require representations that don't have their own classical limit. 
The fact that they cannot be developed by generalizing away from a classical limit seems to impede the full formulation of a final theory even once one has found it. 
The resulting idea of a fundamental theory whose full formulation is hidden from the physicists' grasp because its most adequate representation lacks intuitive roots has even more radical rivals, which amount to questioning the possibility of calculating the dynamics of the fundamental theory either within the bounds of human calculational power or as a matter of principle. 

It has been suggested by various exponents and observers of contemporary fundamental physics (see e.g. Smolin 2003, Woit 2003, Hossenfelder 2018) that the chronic incompleteness of string theory represents a substantial failure of the research program that is indicative of a strategical problem that has afflicted fundamental physics in recent decades.
This conclusion is based on the implicit understanding that the challenges faced by scientific theory building, though obviously differing in their specifics, must always remain of roughly the same general kind. A substantial deviation from expectations regarding the general `performance' of scientific theories in a field on that view must indicate that the way the scientific process plays out in the given case is flawed. 

The present chapter suggests a very different understanding of the current state of physics. Considering the range and character of the very substantial differences that set the current state of fundamental physics apart from any previous stage in the history of physics, there is little reason to expect that theory building at the present stage can be judged according to criteria that seemed adequate in the past.
To be sure, it might still happen that a new conceptual idea will in the near future break the gridlock and, within a moderate timespan, lead up to a complete and predictively powerful theory of quantum gravity. Obviously, none of the arguments presented rules out such a scenario. But, judging the present situation on its own merits rather than based on the experience one has gathered with respect to earlier theories that played out in a very different overall context, that may not be the most plausible scenario.

It may be necessary to substantially correct our understanding of the role of theory building once physics has entered the stage when it plays out in the context of a final theory. It might therefore make sense to significantly downscale expectations on what a theory of quantum gravity can achieve in the foreseeable future. Understanding the character of the suggested shift in detail and analysing which achievements can be reasonably aimed at under the new circumstances will then itself become an important element of physical reasoning.


\section*{References}

\bdes
\item Banks, T., W. Fischler, S.H. Shenker and L. Susskind (1997): "M-theory as a Matrix Model", Physical Review D55: 5112-5128.
\item Cartwright, N. (1983): \textit{How the Laws of Phsics Lie}, Oxford: Clarendon Press.
\item Dawid, R. (2006): "Underdetermination and Theory Succession from the Perspective of String Theory", Philosophy of Science 73/3: 298-322.
\item Dawid, R. (2013): \textit{String Theory and the Scientific Method}, Cambridge: Cambridge University Press.
\item Dawid, R. (2013a): ``Theory Assessment and Final Theory Claim in String Theory", Foundations of Physics 43(1): 81-100, 2013.
\item Dawid, R. (forthcoming): ``The Significance of Non-empirical Confirmation in Fundamental Physics", in \textit{Why Trust a Theory?}, Dardashti, R., R.Dawid and K. Thebault, eds., Cambridge University Press, to appear 2019, PhilSci-Archive/14630.
\item De Haro, S. (2018): ``The Heuristic Function of Duality", PhilSci-Archive/14331.
\item Duff, M. (2015): ``How Fundamental are Fundamental Constants?", Contemporary Physics 56(1): 35-47.
\item Green, M. B. and Schwarz, J. H. (1984): ``Anomaly cancellation in supersymmetric D=10 gauge theory and superstring theory", Phys. Lett. B149, 117-122.
\item Greene, B. (1999): \textit{The Elegant Universe}, Vintage.
\item Hossenfelder, S. (2018): \textit{Lost in Math - How beauty leads Physics Astray}, Basic Books.
\item Maldacena, J. (1998): ``The large N limits of superconformal field theories and supergravity", Adv. Theor.Math. Phys 2: 231 (arXiv:hep-th/9711200).
\item Montonen, C. and D. Olive (1977): ``Magnetic Monopoles as Gauge Particles?", Physics Letters 72B(1): 120.  
\item Polchinski, J. (2017): ``Dualities of Fields and Strings", Studies in the History and Philosophy of Modern Physics 59: S6-20. 
\item Scherk, J. And J. H. Schwarz (1974): Dual Models for Nonhadrons, Nuclear Physics B81, p118.
\item Smolin, L. (2003): \textit{The Trouble with Physics}, Houghton Mifflin.
\item 't Hooft (1993): Dimensional reduction in quantum gravity, Ali, A. and D. Amati (eds.), Salamfestschrift: 284-296, World Scientific.
\item Veneziano, G. (1968): Construction of a Crossing-Symmetric, Regge Behaved Amplitude for Linearly Rising Trajectories, Nuovo Cimento 57A, p190.
\item Witten, E. (1995): ``String theory dynamics in various dimensions", Nucl. Phys B443, 85-126 (arXiv:hep-th/9503124).
\item Witten, E. (1996): Reflections on the Fate of Spacetime, reprinted in Callender, C. and Huggett, N. (eds): \textit{Physics meets Philosophy at the Planck Scale}, Cambridge University Press 2001.
\item Woit (2003): \textit{Not Even Wrong: The Failure of String Theory and the Continuing Challenge to Unify the Laws of Physics}, Jonathan Cape.
\edes

\end{document}